\documentclass[12pt]{article}
\usepackage{amssymb}
\usepackage{amsmath}
\usepackage{apacite}
\usepackage{latexsym}
\usepackage{epsfig}
\usepackage{graphicx}
\usepackage{subfigure}
\usepackage{wasysym}
\usepackage{threeparttable}
\usepackage{natbib}
\usepackage{color}
\usepackage{epstopdf}
\usepackage{bm}
\usepackage{float}
\usepackage{todonotes}
\usepackage{verbatim}

\usepackage{hyperref}

\def\boxit#1{\vbox{\hrule\hbox{\vrule\kern6pt
          \vbox{\kern6pt#1\kern6pt}\kern6pt\vrule}\hrule}}

\def\bse{\begin{eqnarray*}}
\def\ese{\end{eqnarray*}}
\def\be{\begin{eqnarray}}
\def\ee{\end{eqnarray}}
\def\bq{\begin{equation}}
\def\eq{\end{equation}}
\def\bse{\begin{eqnarray*}}
\def\ese{\end{eqnarray*}}

\begin{document}

\thispagestyle{empty} 
\baselineskip=28pt

\begin{center}
{\LARGE{\bf Computationally Efficient Deep Bayesian Unit-Level Modeling of Survey Data under Informative Sampling for Small Area Estimation}}

\end{center}

\baselineskip=12pt

\vskip 2mm
\begin{center}
Paul A. Parker\footnote{(\baselineskip=10pt to whom correspondence should be addressed)
Department of Statistics, University of Missouri,
146 Middlebush Hall, Columbia, MO 65211-6100, paulparker@mail.missouri.edu},
   and Scott H. Holan\footnote{\baselineskip=10pt Department of Statistics, University of Missouri
146 Middlebush Hall, Columbia, MO 65211-6100, holans@missouri.edu}\,\footnote{\baselineskip=10pt U.S. Census Bureau, 4600 Silver Hill Road, Washington, D.C. 20233-9100, scott.holan@census.gov}
\\
\end{center}
%
%
%
%
\vskip 4mm
\baselineskip=12pt 
\begin{center}
{\bf Abstract}
\end{center}

The topic of deep learning has seen a surge of interest in recent years both within and outside of the field of Statistics. Deep models leverage both nonlinearity and interaction effects to provide superior predictions in many cases when compared to linear or generalized linear models. However, one of the main challenges with deep modeling approaches is quantification of uncertainty. The use of random weight models, such as the popularized “Extreme Learning Machine,” offer a potential solution in this regard. In addition to uncertainty quantification, these models are extremely computationally efficient as they do not require optimization through stochastic gradient descent, which is what is typically done for deep learning. We show how the use of random weights in a deep model can fit into a likelihood based framework to allow for uncertainty quantification of the model parameters and any desired estimates. Furthermore, we show how this approach can be used to account for informative sampling of survey data through the use of a pseudo-likelihood. We illustrate the effectiveness of this methodology through simulation and with a real survey data application involving American National Election Studies data.
\baselineskip=12pt
\par\vfill\noindent
\par\medskip\noindent
\clearpage\pagebreak\newpage \pagenumbering{arabic}
\baselineskip=12pt

	\section{Introduction}
	
	There has recently been a strong interest in collecting novel types of data along with sample surveys. These data types can range from functional data such as the physical activity monitor data contained within the National Health and Nutrition Examination Survey \citep[NHANES;][]{schuna2013adult}, free response text data such as those within the American National Election Studies (ANES) surveys \citep{debell2013harder}, and even complex data from web-based surveys such as mouse movements \citep{horwitz2020learning}. These information rich data sources could prove to be useful as covariates, however the rapid and increased interest in these data types within a survey context has resulted in lagging development of corresponding methodology.
	
	These complex covariates are generally collected at the unit level, and thus, necessitate the need for unit-level modeling strategies. One of the challenges associated with unit-level modeling is accounting for the sampling design under informative sampling mechanisms. A variety of strategies exist for this problem, including the use of pseudo-likelihood modeling \citep{ski89, bin83}. In many applications, such as small area estimation, predictions can be made for every unit in the population and then aggregated as necessary to construct any desired estimates. An exceedingly common solution is that of regression and poststratification, whereby the population is segmented via a set of categorical covariates, and units that are associated with identical covariates are assumed to be independent and identically distributed \citep{park2004bayesian}. In particular, \citet{park2004bayesian} state that they envision this methodology being used for public opinion estimates at the state level. The categorical covariates required for poststratification are generally known for the entire population. This is in contrast to the complex covariates that we consider, which are generally only known for the sampled units.
	
	Another major concern for unit-level models, particularly in a Bayesian setting, is that of computational efficiency. Dependent data models typically rely on Gaussian prior distributions for model parameters \citep{brad15}; however, most survey variables tend to be non-Gaussian, leading to non-conjugate conditional distributions that can be difficult to sample from. This problem is addressed by \cite{parker20} and \citet{parker20b} for count data and Binomial data, although they do not consider the further problem of modeling complex data types in a computationally efficient manner.
	
	Herein, we develop a computationally efficient method to model these complex covariates while accounting for informative sampling. Although other applications of this model are possible, we illustrate this method through the problem of small area estimation. We utilize a neural network structure to handle nonlinear modeling of the complex covariate data, while employing a Bayesian pseudo-likelihood model structure in order to measure uncertainty around our estimates while accounting for informative sampling. The remainder of this paper is outlined as follows. In Section \ref{sec: methods} we introduce the necessary methodological background as well as our model. Section \ref{sec: sim} considers an empirical simulation study relying on the use of ANES data. We follow with a full data analysis using the same ANES data in Section \ref{sec: DA}. Finally, we provide discussion and concluding remarks in Section \ref{sec: disc}.

	\section{Methodology}\label{sec: methods}
	
	The method that we develop tackles three problems simultaneously. The first issue is that nonlinear modeling is required for the use of complex covariates, without becoming computationally prohibitive. For example, neural network structures typically involve an extremely high-dimensional parameter space. A full Bayesian treatment of these types of models can often require too many computational resources to be fit in a reasonable amount of time. The second problem is that we must account for informative sampling in our model to avoid producing any unnecessary statistical bias. Finally, we require a model that can handle Binomial data types while still accounting for all of the underlying dependencies associated with the data. We explore each of these three problems, and then present our methodology.
	
	\subsection{Extreme Learning Machines}
	
	The extreme learning machine (ELM) is a type of single layer feed-forward neural network (FNN), introduced by \citet{huang06}. The key difference between the ELM and traditional FNNs is that the ELM uses random weights (i.e., parameters) drawn from some distribution for the hidden layer nodes. As with other FNNs, the ELM can be used for both regression and classification problems, as well as other types of problems (e.g., unsupervised learning), while allowing for much more flexibility in the mean function than linear or generalized linear models.

The basic ELM considers a nonlinear transformation of the covariate data (features),
$$
f(\mathbf{x}_i) = \sum_{j=1}^N g_j(\mathbf{a}_j'\mathbf{x}_i + b_j)\beta_j, \; i=1,\ldots,n
$$ where $\mathbf{x}_i$ represents the $p$-dimensional covariate information for unit $i$ in the sample with size $n$. The value $N$ represents the number of nodes, where each node considers a unique nonlinear transformation of the data. Each node first applies a linear transformation, with parameters $\mathbf{a}_j=[a_{j1},\ldots,a_{jp}]'$ and $b_j$. This linear transformation is followed by a nonlinear transformation, denoted by the function $g_j(\cdot)$, often called an activation function. This is specified a priori, and may be any piecewise continuous function \citep{huang15}, but in practice usually consists of a sigmoid function. The output, $f(\mathbf{x}_i)$ is then calculated as a weighted sum of each individual node output, where the weights are denoted by the $N$ dimensional vector $\boldsymbol{\beta}$. Intuitively, this is similar to basis function approaches, in the sense that both techniques may use nonlinear transformations of the input data to construct a flexible nonlinear function for the mean. However, one key difference is that in the case of the ELM, these nonlinear transformations do not need to be selected as they are randomly generated.

The key to ELMs is that for each node, $j=1,\ldots,N$, the values of $\mathbf{a}_j$ and $b_j$ are randomly drawn. Thus, the only set of parameters that need to be learned or estimated is $\boldsymbol{\beta}$. Common distributional choices for these randomly selected parameters are Normal(0,1) and Uniform(-1,1). Although these hidden layer parameters are only randomly drawn a single time, typically many nodes are used to allow for flexible representation of the function $f(\mathbf{x}_i)$.

More generally, the ELM can be written
$$
\begin{aligned}
  \bm{\mu}_i &= g_o(\mathbf{Bg}_i) \\
  \mathbf{g}_i &= g(\mathbf{Ax}_i)
\end{aligned}
$$
where the $p \times 1$ dimensional vector $\mathbf{x}_i$ now contains an intercept and the $l$-dimensional vector of means, $\bm{\mu}_i$, can now incorporate multivariate responses. Also, $\mathbf{A}$ is an $N \times (p+1)$ dimensional matrix of hidden layer weights, and $\mathbf{B}$ is an $l \times N$ dimensional vector of output weights. The hidden layer activation function is denoted $g(\cdot)$ and the output layer activation function is denoted $g_o(\cdot)$, which will be the inverse of the canonical link function in the case of a GLM. 

The above view of the ELM is similar to the generalized linear model and highlights an important strength of ELM. Because the hidden layer parameters are randomly chosen and not estimated, we may view the hidden layer transformations as fixed once these parameters have been generated. This is similar to regression with basis expansions as is often seen when using generalized additive models (GAMs). A key difference with ELM compared to GAMs however, is that the entire vector $\mathbf{x}_i$ is used within each hidden node, which allows for interaction effects. Viewing the random transformations as fixed allows us to extend the entire class of generalized linear models to incorporate nonlinear behavior. Furthermore, pseudo-likelihood approaches may be used in conjunction with the ELM in order to account for informative sampling.

The ELM can be considered a type of reservoir computing (an approach where weights are randomly generated). Random projection is another type of reservoir computing, often used for dimension reduction \citep{bingham01}. Under random projection, the original $n\times p$ data matrix $\mathbf{X}$ is ``projected" onto a $L$-dimensional subspace, $\mathbf{X}^*=\mathbf{XR}$, where $\mathbf{R}$ is a randomly generated $p \times L$ matrix. This is not technically a projection, as the matrix $\mathbf{R}$ is not orthogonal, but due to the random nature of the matrix, it tends to be ``approximately" orthogonal. Note that the randomly generated projection matrix could be orthogonalized, but this is not always done in practice due to the computational cost.

Random projection could be used in the context of regression, similar to Principal Components Regression. In this light, it may be seen as a special case of the ELM, where $g_j(\cdot)$ is equal to the identity function for all $j$. In other words, random projection uses randomly generated parameters for the hidden node linear transformation component, but does not introduce a nonlinear component.

Another common type of reservoir computing is known as the Echo State Network or ESN \citep{prok05}. This is a type of recurrent neural network, where the hidden weights are randomly generated. Recently, the ESN has been used in likelihood-based frameworks for spatio-temporal forecasting \citep{mcdermott17}. The ESN may also be used within a Bayesian model structure in order to give uncertainty quantification \citep{mcdermott19}.

Bayesian methods have been considered in the ELM community as well, beginning with \citet{soria11}. They consider ridge regression fit with an Empirical Bayes procedure. This achieves both regularization as well as uncertainty quantification for the output layer weights and data model variance. They also show that this method tends to give better out of sample predictions compared to the traditional ELM. \citet{chen16} use a variational Bayes approach to fit a Bayesian ELM. By doing so, it is possible to reduce the computational burden of the Bayesian ELM substantially.
	
	\subsection{Pseudo-likelihood based SAE}
	
	When fitting models with unit-level survey data, it may be the case that there exists a dependence relationship between the unit probabilities of selection, and the response values. This is termed {\it informative sampling}, and if this relationship is not accounted for, any estimates may be biased \citep{pfe07}. A thorough review of the modern approaches to handling informative sampling is given by \citet{parker2019unit}. One popular approach to this problem is the use of a pseudo-likelihood (PL), introduced by \citet{ski89} and \citet{bin83}. The general idea is to use the reported survey weights to exponentially re-weight the likelihood contribution of each survey unit. Thus, the PL is written as
	\begin{equation}\label{E: pseudoLikelihood}
  \prod_{i \in \mathcal{S}}  f( y_i \mid \boldsymbol{\theta})^{w_i},
\end{equation} where $y_i$ is the response value and $w_i$ is the survey weight for unit $i$ in the sample $\mathcal{S}$. The PL can be maximized in order to make frequentist inference, however \citet{sav16} show that in a Bayesian setting, the pseudo-posterior distribution,
\begin{eqnarray*}
\hat{\pi}(\bm{\theta} | \mathbf{y}, \mathbf{\tilde{w}}) \propto \left\{ \prod_{i \in \mathcal{S}} f(y_{i} | \bm{\theta})^{\tilde{w}_{i}} \right\} \pi (\bm{\theta}),
\end{eqnarray*} converges to the population posterior distribution, justifying the use of a Bayesian PL for inference on nonsampled units. In this case, $\tilde{w_i}$ represents the survey weights after scaling to sum to the sample size in order to give proper uncertainty quantification.

	\subsection{Logistic Models}
	
	Many survey data variables tend to be non-Gaussian at the unit level. For example, the American Community Survey contains a binary indicator of health insurance status as well as many categorical variables such as primary language spoken. In regression frameworks with non-Gaussian responses and Normal prior distributions on any regression parameters, non-conjugate full conditional distributions arise. This may lead to the need for Metropolis steps within the MCMC routine that can be prohibitively difficult to tune.
	
	For the case of logistic models (Binomial, Negative Binomial and Multinomial responses), \citet{pol13} introduce a data augmentation scheme that gives rise to conjugate full-conditional distributions. This strategy relies on the use of P\'{o}lya-Gamma (PG) random variables.  Specifically, they rewrite the Binomial likelihood as,
	\begin{equation}\label{eq: pg}
   \frac{(e^{\psi})^a}{(1 + e^{\psi})^b} = 2^{-b}e^{\kappa \psi} \int_0^{\infty} e^{-\omega \psi^2/2} p(\omega) d\omega, 
\end{equation} where $\kappa = a - b/2$ and $p(\omega)$ is a $\hbox{PG}(b,0)$ density. They also show that $p(\omega | \psi) \sim \hbox{PG}(b,\psi)$. For the linear predictor $\psi=\bm{x}'\bm{\beta}$, if we use a Gaussian prior on $\bm{\beta}$, the full conditional distribution for $\bm{\beta}$ will also be Gaussian. Furthermore, \citet{parker20b} show that under a PL setup, conjugacy is still retained. They develop both a Gibbs sampling algorithm as well as a variational Bayes algorithm for PL-based mixed effects models with Binomial data. In addition to this, they use the stick-breaking representation of the Multinomial distribution in order to extend the algorithms to categorical responses. More specifically, \citet{linderman15} show that the Multinomial distribution may be written as a product of independent Binomial distributions,
\begin{equation}
    \begin{split}
        \hbox{Multinomial}(\bm{Z}|n, \bm{p})=\prod_{k=1}^{K-1} \hbox{Bin}(Z_k|n_k, \tilde{p}_k),
    \end{split}
\end{equation} where
\begin{equation}
    n_k = n - \sum_{j < k}x_j, \; \; \tilde{p}_k = \frac{p_k}{1 - \sum_{j<k} p_j}, \; \; k=2,\ldots,K.
\end{equation} Under this view of Multinomial data,  $K-1$ Binomial data models may be fit independently while still accounting for the dependence between categories through the stick-breaking counts and probabilities.
	
	\subsection{Proposed Model}
	
	We now introduce a Bayesian unit-level deep model for informative sampling (BUDIS). Here, we focus on the case of Binomial and Multinomial data, but note that this approach would be applicable to Gaussian data as well. The Binomial model is written,
	    \begin{equation}
    \begin{split}
        \bm{Z} | \bm{\beta, \eta} & \propto \prod_{i \in S} \left\{\hbox{Bin}\left(Z_i | n_i, p_i \right)^{\stackrel{\sim}{w}_i}\right\} \\
        \hbox{logit}(p_i) &= \bm{x}_i'\bm{\beta} +  \bm{g}_i' \bm{\eta}  \\
        \bm{g_i'} &= \frac{1}{1 + e^{\bm{-A \psi_i}}} \\
        \bm{\eta}|\sigma^2_{\eta} & \sim \hbox{N}_h(\bm{0}_h, \sigma_{\eta}^2 \bm{I}_h ) \\
        \bm{\beta} & \sim \hbox{N}_p(\bm{0}_p, \sigma_{\beta}^2 \bm{I}_p ) \\
        \sigma_{\eta}^2 & \sim \hbox{IG}(a, b) \\
        & a, b, \sigma^2_{\beta} >0,
    \end{split} 
\end{equation} where $Z_i$ is the Binomial response for unit $i$ in the sample with size $n_i$ and probability $p_i$. Typically surveys contain Bernoulli data, so for our purposes, $n_i=1$ for all $i$. We are using a pseudo-likelihood approach at this data stage of the model in order to account for informative sampling. Within the pseudo-likelihood, we use the scaled survey weights, $\tilde{w}_i$, such that the weight sum to the sample size. The length $p$ vector $\bm{x}_i$ contains any covariates that do not require nonlinear modeling. The length $h$ vector $\bm{g}_i$ contains the ELM hidden layer values for unit $i$. Finally, the length $r$ vector $\bm{\psi}_i$ contains the complex covariates that are used within the ELM framework.  Note that the $h \times r$ matrix $\bm{A}$ is sampled and considered fixed before model fitting, so that $\bm{g}_i$ is determined {\it a priori}. This allows for the use of generlized linear model fitting procedures rather than custom techniques. Specifically, we use the variational Bayes procedure from \citet{parker20b} for all model fitting.

For our purposes, we let $\bm{x}_i$ consist of any poststratification variables as well as spatial basis functions. These values will typically be known for the full population. The complex covariates contained within $\bm{\psi}_i$ are not usually known for the full population, and thus must be imputed in order to generate the population posterior predictive distribution necessary for small area estimation. Our approach revolves around the idea of assigning the observed covariate vectors to all the unobserved population units. Under a simple random sample, a reasonable assumption may be that the observed complex covariates are uniformly distributed throughout the population. However, under an informative sample, the observed covariate vectors are sampled with unequal probability, which should be accounted for when distributing the observed vectors to the population.

For our imputation model, we create imputation cells, similar to poststratification cells, with $J$ total cells. A population unit within imputation cell $j, \; j=1,\ldots,J$ may only be assigned a vector from the set of observed vectors $(\bm{\psi}_{j1},\ldots, \bm{\psi}_{jn_j})$, where $n_j$ is the sample size within cell $j$. Rather than sampling from this set with equal probability, we sample with probability proportional to the reported sampling probability, or inversely proportional to the reported sample weight, to account for the original survey sampling scheme. Thus, for population unit $i$ in cell $j$, the vector of complex covariates is sampled from $(\bm{\psi}_{j1},\ldots, \bm{\psi}_{jn_j})$ with probability proportional to $(1/w_{j1},\ldots, 1/w_{jn_j})$. This imputation can be done a single time, however we opt to create a separate imputed dataset for each sample from our model based posterior distribution in order to account for the imputation uncertainty within our posterior predictive distribution. For this work, we let the $J=48$ corresponding to the states where area level estimates are made, however other choices of imputation cells could be explored. One limitation to this approach is that all imputation cells must have at least one sample unit in order to distribute the sample values within the cell to the population.

	\section{Empirical Simulation Study}\label{sec: sim}
	
	To test our methodology, we consider data from the 2012 American National Election Studies (ANES) survey. Specifically, we use the Time Series Study data which measures various responses both pre and post election. We only consider the post election data, which contains a number of free response questions. Our goal is to use the free responses from the question ``What are the most important problems facing this country?" in order to improve small area estimates of public opinion.
	
\begin{figure}[H]
                \begin{center}
                \includegraphics[width=150mm]{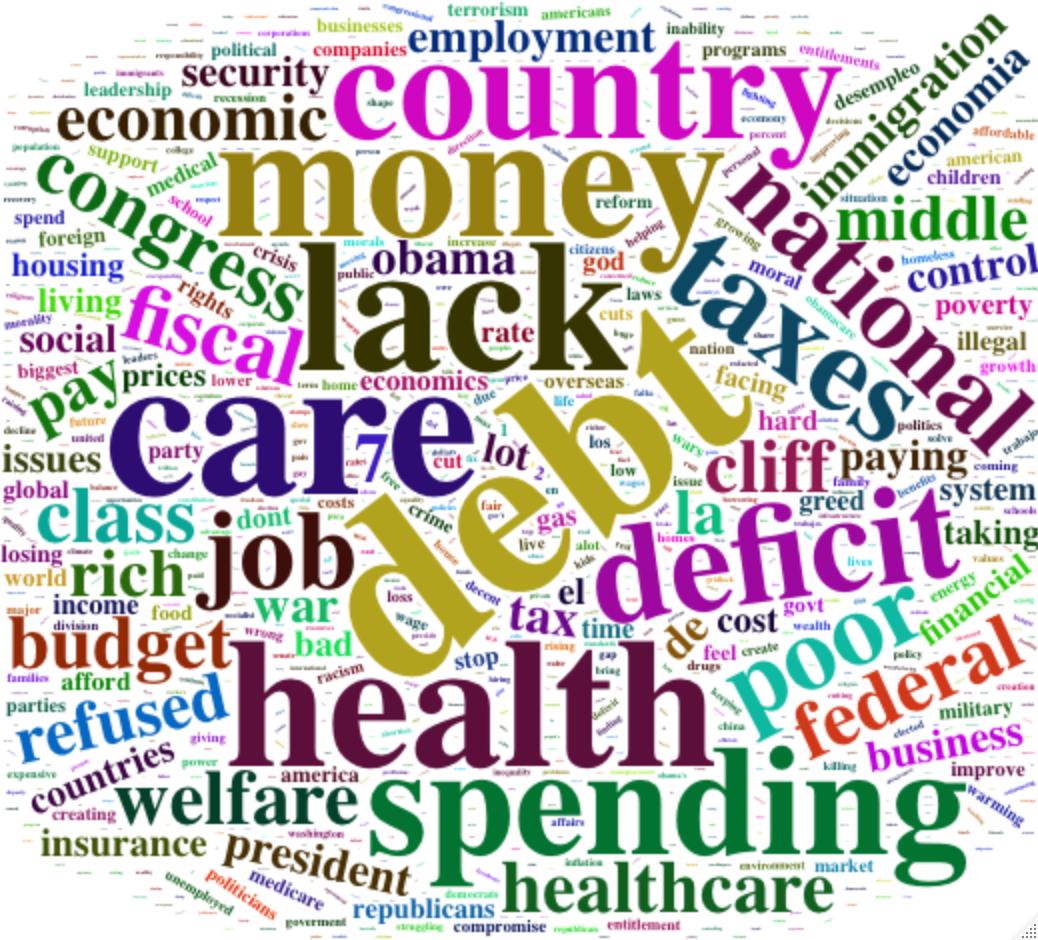}
                 \caption{\baselineskip=10pt A word cloud of the top words contained within the ANES data. The size of each word represents the frequency of appearance.}
                 \label{fig:words}
 \end{center}               

            \end{figure}

	Figure \ref{fig:words} shows a word cloud of the most frequently occurring words within the ANES data. In many cases, the words will have little meaning on their own, but instead have meaning when paired with other words. For example, the word {\it security} on its own does not provide much insight, but when paired with either {\it economic} or {\it national}, it may indicate the primary concern of the respondent. This suggests the need for a model that can take into account many possible interactions between words rather than considering words individually.
	
	One public opinion question involved in the survey considers whether respondents approve or dissaprove of the way president Barack Obama was handling the job of president at the time. For this simulation, we estimate the proportion of the population within each state that approve. In other words, we consider a binary response. We treat the original ANES sample as our population and take a subsample with probability to proportional size sampling using the Poisson method \citep{brewer1984poisson} with an expected sample size of 1,000. For our size variable, we use the original survey weight plus 0.7 if the true response is ``approve" in order to explicitly generate an informative sample. We fit the BUDIS model using $\bm{\psi}_i=(I(w_{i1}),\ldots,I(w_{i1000}))$ as the input into the ELM, where $I(w_{ij})$ indicates whether or not the $j$th most frequently occurring word appeared in the free response of unit $i$. For the linear component, $\bm{x}_i$, we use indicators of Hispanic ethnicity and gender as poststratification variables, as well as a set of spatial basis functions. We use the first 25 eigenvectors of the state adjacency matrix as our basis functions, although other basis functions could be substituted here. We generate our hidden weights in the matrix $\bm{A}$ from the standard Normal distribution, and then randomly set 10\% equal to zero and we use a vague prior distribution by setting $a=b=0.5$ and $\sigma^2_{\beta}=1000$. We have found that in general the model is not overly sensitive to the choice of distribution for the random weights (i.e. the generating distribution for the matrix $\bm{A}$), however, depending on the application, it may be desirable to select the distribution through cross-validation. Lastly, we set the number of hidden nodes $h=240$. We also compare to a model that does not use the text data or the ELM component, which we denote Pseudo-likelihood logistic regression (PLLR), as well as both weighted and unweighted direct estimators. The two model based approaches both use post-stratification by sampling from the posterior distribution of the parameters, and then generating estimates for the response value of all units in the population. We repeat the sampling and model fitting procedure 50 times.
	
	Table \ref{table:1} shows the MSE and squared bias of each of the estimators through the simulation. The first thing to note is that the much higher bias of the unweighted (UW) direct estimator when compared to the direct estimator indicates that the sampling was indeed informative. The two model based approaches were able to handle the informative sampling mechanism through the use of the pseudo-likelihood and improve the MSE dramatically compared to the direct estimates. The BUDIS model was able to further improve upon the PLLR model by reducing MSE about 10\% and reducing squared bias around 21\%. It is clear in this case that the inclusion of nonlinear modeling of the text covariates results in better estimates.
	
\begin{table}[H]
\begin{center}
 \begin{tabular}{||c | c c ||} 
 \hline
 Estimator & MSE & Bias$^2$  \\ [0.5ex] 
 \hline\hline
 BUDIS & $\bm{1.99 \times 10^{-2}}$ & $6.43 \times 10^{-3}$  \\ 
 \hline
 PLLR & $2.21 \times 10^{-2}$ & $8.09 \times 10^{-3}$  \\
 \hline
 Direct & $4.49 \times 10^{-2}$ & $\bm{4.72 \times 10^{-3}}$  \\
 \hline
 UW Direct & $4.21 \times 10^{-2}$ & $1.33 \times 10^{-2}$  \\ [1ex] 
 \hline
\end{tabular}
\caption{MSE and squared bias of the four estimators based on simulation results}
\label{table:1}
\end{center}
\end{table}

	\section{ANES Data Analysis}\label{sec: DA}
	In order to illustrate this methodology on a real application, we use the entire 2012 ANES dataset to create estimates under the BUDIS model. The total sample size for this dataset was 5,878, with state sample sizes ranging from 4 (Wyoming) to 742 (California). Similar to the simulation study considered in Section \ref{sec: sim}, we estimate the proportion of voting age residents within each state that approved of Barack Obama's job as president at the time the survey was taken. The covariates and hyperparameters were also the same as those considered in the simulation study.
	
\begin{figure}[H]
                \begin{center}
                \includegraphics[width=150mm]{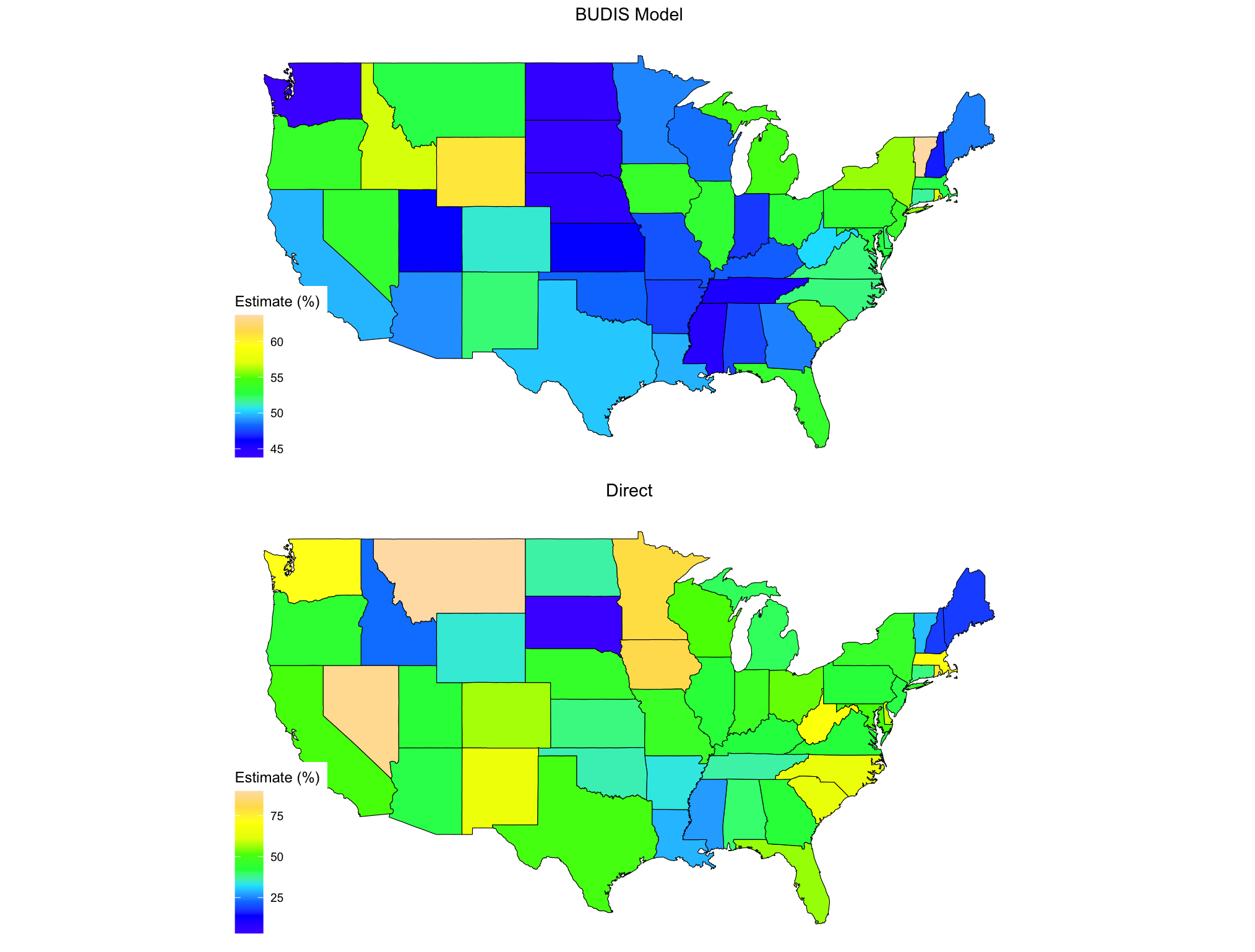}
                 \caption{\baselineskip=10pt Comparison of BUDIS model and direct estimates on 2012 ANES data. Note that separate scales are used in order to emphasize the spatial pattern under each approach.}
                 \label{fig:map}
 \end{center}   \end{figure}

	We compare the estimates under the BUDIS model to the direct estimates in Figure \ref{fig:map}. Note that many of the direct estimates fall towards the extremes due to limited sample sizes in some states. In contrast to this, the model based estimates fall in a narrower range due the effect of ``borrowing information" across states. For the most part the spatial pattern under the BUDIS model is as expected. The more traditionally conservative states in the South and towards the Dakotas tend to have lower estimates of approval than the coastal parts of the country. The Northwest portion of the country has a couple unexpected estimates, namely Wyoming and Washington. The higher than expected estimate for Wyoming is likely due to the limited sample size pulling the estimate upward towards the national average, although an effort to find more suitable spatial basis functions could also aid improvement in this area.
	
 This example emphasizes how the BUDIS model can be used to construct better estimates of public opinion through the use of complex data types such as free response text. The ANES data set has a sufficient sample size to construct direct estimates at the national level, but many of the state sample sizes are extremely small, leading to very poor estimates. In this specific case, many of the state level direct estimates fall far away from from the national average. In contrast, the BUDIS model is able to smooth many of these extreme estimates by relying on a model that accounts for spatial dependence between survey respondents as well as the complex free response covariate information. Despite the small state sample sizes, this model is able to yield much more reasonable estimates, such as the general trend of lower approval in the South and higher approval on the East coast. These types of estimates could serve useful for targeting of campaign funds. For instance, in the key states of Florida and Michigan, the model based estimates indicate the Michigan may be more competitive. Furthermore, this example only considers estimates for a single public opinion question, but the ANES survey contains many more public opinion questions that may be of interest to others.
 
	\section{Discussion}\label{sec: disc}
	
In order to use complex unit level survey data as a covariate for small area estimation, we develop a couple important innovations to the PL unit-level model. The first innovation is the use of deep learning to model nonlinear functions of the complex covariates. This is achieved through the use of random weight methodologies, specifically the ELM. By taking this approach we are able to side-step the need for gradient descent techniques that are typically used in deep learning. In addition, this approach is highly computationally efficient, as it is linear in the parameters that are estimated. Further efficiency is gained through the use of a variational Bayes model fitting procedure. 

The second innovation is the use of an imputation model that assigns sample covariates to population units while adjusting for the sampling design. Although this approach is relatively straightforward, modeling the population covariates explicitly could be very burdensome for high-dimensional data and this approach provides a path forward. The use of more advanced approaches to this imputation problem is subject to future work.

In addition to the novel modeling approaches explored here, this work highlights the need to collect more complex data types within surveys. Typical surveys include relatively simple data types such as binary and categorical measurements. However this work shows that more complex data types such as text or functional data may be used to improve the precision of survey based estimates. Currently, federal agencies spend significant resources converting open responses into simple categorical variables. Through the use of our proposed model, or extensions thereof, agencies may be able to rely less on these resources while simultaneously extracting more information from the raw data.  The ANES data considered in our examples was chosen in part because of its public availability, in order to limit the need for disclosure issues. However, our approach could be immediately (or with minor modifications) applicable to other complex survey datasets, such as NHANES physical activity monitor data, or web-based respondent tracking.

\section*{Acknowledgements}
Support for this research through the Census Bureau Dissertation Fellowship program is gratefully acknowledged.  This research was partially supported by the U.S.~National Science Foundation (NSF) under NSF grant SES-1853096. This article is released to inform interested parties of ongoing research and to encourage discussion. The views expressed on statistical issues are those of the authors and not those of the NSF or U.S. Census Bureau.

\bibliography{elmBib}
\bibliographystyle{jasa}

\end{document}